\documentstyle[pra,epsf,aps,twocolumn]{revtex}

\def\lfrac#1#2{#1/#2}
\def\db{\,\, {\bar{} \!\!d}\!\,\hspace{0,5pt}}
\def\comment#1{}
\def\lfrac#1#2{#1/#2}

\begin{document}
\sloppy
\title{Reparametrization Invariance of Perturbatively Defined
  Path Integrals.\\
II. Integrating Products of Distributions}
\author{H.~Kleinert\thanks{E-mail: kleinert@physik.fu-berlin.de} and
     A.~Chervyakov\thanks{On leave from LCTA, JINR, Dubna, Russia
                   E-mail: chervyak@physik.fu-berlin.de}
                      \\ Freie Universit\"at Berlin\\
          Institut f\"ur Theoretische Physik\\
          Arnimallee14, D-14195 Berlin}
\maketitle
\begin{abstract}
We show how to
perform  integrals  over products of
distributions
in coordinate space such as to reproduce
the results
of momentum space Feynman integrals
in dimensional regularization.
This ensures the invariance
of path integrals under coordinate transformations.
The integrals are performed
by
 expressing the
propagators in $1- \varepsilon $ dimensions
in terms of  modified Bessel functions.
\end{abstract}
\section{Introduction}
While the Schr\"odinger equation and
time-sliced path integrals, as defined in
the textbook
\cite{1}, are invariant under coordinate transformations by construction,
this invariance is completely nontrivial
in
perturbatively defined path integrals
in curvilinear coordinates.
The reason for this is that
the
Feynman integrals
appearing in the perturbation
expansion are multiple temporal integrals
over highly singular
integrands,  which are products of generalized functions
[for general discussion, see \cite{qft} ].
In the presently existing theory of distributions,
generalized functions can only be combined linearly and by a convolution,
but products of them are still forbidden calling for
regularization \cite{gf}.
Thus, if we want to describe quantum mechanics
in curvilinear coordinates by perturbative defined path integrals,
we must extend the theory of distributions
to allow for integrals over products of generalized functions.
This will extend the
linear space of distributions to a group.

In a previous note \cite{2} we have
shown
that invariance can be achieved
by defining path integrals perturbatively
as a limit $D\rightarrow 1$ of a $D$-dimensional
functional integral.
The perturbative calculations
in \cite{2}
were performed in momentum space,
where
Feynman integrals
in
a continuous number of dimensions
$D$ are known from the prescriptions of
't~Hooft and M.~Veltman\cite{3}.
For many applications, however,
momentum space calculations are rather clumsy,
for instance if one wants to find the effective
action of a field system in curvilinear
coordinates, where the kinetic term depends
on the dynamic variable.
Then one
needs rules for performing temporal integrals
over
Wick contractions
of local fields.

\section{Problem with Coordinate Transformations}
Recall the origin of the difficulties
with coordinate transformations
in path integrals.
Let $q(\tau )$ be the euclidean coordinates of a quantum-mechanical
point particle of unit mass in a harmonic potential $m^2q^2/2$
as a function of the imaginary time $\tau =-it$.
Under a coordinate transformation
 $q(\tau)\rightarrow \phi(\tau )$
defined by $q (\tau )=f(\phi(\tau ))=\phi(\tau )+\sum _{n=1}^\infty a_n
\phi^n(\tau )$,
the kinetic term $\dot q^2(\tau )/2$
goes over into $\dot \phi^2(\tau )f'{}^2(\phi(\tau ))/2$.
If the path integral over $\phi(\tau )$ is performed
 perturbatively, the
expansion terms
contains temporal integrals
over
Wick
contractions which,
after suitable partial integrations,
are products of the following basic correlation functions
\begin{eqnarray}
\Delta (\tau -\tau ')&\equiv& \langle \phi(\tau )\phi(\tau ')\rangle=
\hspace{0mm}\raisebox{-1mm}{\mbox{\input 1.tps }} ,~\label{@D1}\\
\partial _\tau \Delta(\tau -\tau ')&\equiv &\langle \dot \phi(\tau )\phi(\tau ')\rangle
=\hspace{0mm}\raisebox{-1mm}{\mbox{\input 3.tps }} ,~\label{@D2}\\
 \partial_ \tau \partial_{\tau '}\Delta (\tau-\tau ')&\equiv& \langle \dot \phi(\tau )
  \dot\phi(\tau ')\rangle
=\hspace{0mm}\raisebox{-1mm}{\mbox{\input 2.tps }}.~\label{@D3}
\label{@}\end{eqnarray}
The right-hand sides define the line symbols
to be used in Feynman diagrams for the interaction terms.

Explictly,
the first correlation function (\ref{@D1}) reads
\begin{equation}
 \Delta (\tau ,\tau ')=\frac{1}{2m}e^{-m|\tau -\tau '|}.
\label{@del1}\end{equation}
The second correlation function (\ref{@D2})
has a discontinuity
\begin{equation}
\partial _\tau \Delta(\tau ,\tau ') =
     - \frac{1}{2} \epsilon (\tau - \tau ') e^{-m|\tau -\tau '|} ,
\label{@del2}\end{equation}
where
\begin{equation}
\epsilon (\tau - \tau ')\equiv 2\int_{-\infty}^\tau  d\tau''  \delta (\tau'' -\tau ')
\label{@}\end{equation}
is a distribution
which has a
jump at $\tau =\tau '$.
The third correlation function (\ref{@D3}) contains a
$ \delta $-function:
\begin{equation}
 \partial_ \tau \partial_{\tau '}\Delta (\tau, \tau ') =
  \delta(\tau -\tau ') - \frac{m}{2}e^{-m|\tau -\tau '|} ,
\label{@del3}\end{equation}
The temporal
integrals in $\tau $-space over products of such distributions
are undefined.

In our previous paper \cite{2}
we have shown
that it is possible
to define
unique perturbation expansions which lead
to a
reparametrization invariant theory.
This is done by extending the path integral to a $D$-dimensional
functional integral, and by performing the
perturbation
expansion in $D$-dimensional momentum space,
with a limit $D\rightarrow 1$ taken at the end.

In this note we want to set up
an extension of distribution theory
which allows us
to define the same theory
by doing the integrals over products
of the correlation functions
in $\tau $-space. This requires  regularizing the
correlation functions in such a way that all $\tau $-integrals
yield the same results as the momentum integrals
in the limit $D\rightarrow 1$.
The result is
a reparametrization invariant
perturbative definition
of path integrals
via Feynman integrals in $\tau $-space.

\section{Model System}
As in our momentum space treatment in Ref.~\cite{2},
 we  shall prove the reparametrization invariance
for a simple but typical
system, consisting of a point particle in a box of size $d$.
We shall do this by
performing
explicit calculation
up to
three loops.

The ground-state energy of this system is exactly known,
$E =  \pi ^2 /2d^2$. In a recent paper we have shown that this energy
can be obtained perturbatively
to any desired accuracy by performing perturbation
expansions in powers of $g\equiv \pi ^2/d^2 $,
and taking the limit  $g\rightarrow \infty$
of this series \cite{pert},
following the strong-coupling theory
developed in Ref.~\cite{SC}.

In the model system,  the three-loop expansion
for the ground state energy reads
\begin{equation}
 E = \frac{m}{2} + \frac{g}{4} + \frac{1}{16} \frac{g^2}{m} +
   {\cal O}(g^3 ),
\label{exp1}\end{equation}
where $m$ regularizes the Feynman integral
in the infrared. The
ground state energy is obtained from
the limit
$m\rightarrow~\infty$.

After the coordinate transformation of the free-particle action
in the path integral  by
$q (\tau )=f(\phi(\tau )) = \phi - {g}\phi^3/3 + {g^2}\phi^5/5 - \cdots~$,
we find for the energy
the
following graphical expansion \cite{2}:
\begin{eqnarray}
E&=&
\frac{m}{2}
-g\hspace{0mm}\raisebox{-1mm}{\mbox{\input 6.tps }} +\frac{9}{2}\,g^2\hspace{0mm}\raisebox{-3.2mm}{\mbox{\input 7.tps }}
\nonumber \\&&
-\frac{g^2}{2!}\left[~~ 4\hspace{0mm}\raisebox{-1.2mm}{\mbox{\input 8.tps }}
+2\hspace{0mm}\raisebox{-1.2mm}{\mbox{\input 9.tps }}
+2\hspace{0mm}\raisebox{-1.2mm}{\mbox{\input 10.tps }}
\right.\nonumber \\
&&\left.~~~~~~~+4\hspace{0mm}\raisebox{-2mm}{\mbox{\input 11.tps }}
+16\hspace{0mm}\raisebox{-1.95mm}{\mbox{\input 12.tps }}
+4\hspace{0mm}\raisebox{-1.9mm}{\mbox{\input 13.tps }} \right] .
\label{Fig 2}\end{eqnarray}
The path
integral is extended
to an associated functional integral
in a $D$-dimensional coordinate space $x$,
with coordinates $x_\mu\equiv (\tau ,x_2,x_3,\dots)$,
by replacing $\dot\phi^2(\tau )$ in the kinetic term
by $(\partial_\mu\phi(x))^2$,
where $\partial _\mu=\partial /\partial x_\mu$.
In the extended coordinate space, the correlation function
(\ref{@del1})
becomes
\begin{equation}
\Delta (x) = \int \frac{d^Dk}{(2 \pi )^D} \frac{e^{ikx}} {k^2 + m^2},
\label{1}\end{equation}

  In writing down the expansion (\ref{Fig 2})
we have ignored all diagrams coming from
the Jacobian of the coordinate transformation,
since these carry a prefactor
$ \delta^{(D)} (0)$ which,
according to a basic rule
of t'Hooft and Veltman \cite{3},  vanishes
in dimensional regularization.

In our calculations, we shall encounter
generalized $ \delta $-functions, which are multiple derivatives of
the ordinary $ \delta $-function:
\begin{eqnarray}
  \delta ^{(D)} _{\mu_1 \dots \mu_n} (x)& \equiv&
\partial_{\mu_1 \dots \mu_n}  \delta ^{(D)}  (x)\nonumber \\
& =&
\int \db^D k
(ik)_{\mu_1} \dots (ik)_{ \mu_n} e^{ikx},
\label{10}\end{eqnarray}
with
$\partial_{\mu_1 \dots \mu_n} \equiv
\partial_{\mu_1} \dots \partial _{\mu_n}$,
and
with $\db^D k \equiv d^D k \big/ (2 \pi )^D$.
In dimensional regularization,
these vanish at the origin
as well:
\begin{equation}
   \delta  _{{\mu_1 }\dots \mu_n}^{(D)} (0) = \int \db^D k (i k)_{\mu_1}
 \dots (i k)_{\mu_n} = 0 ,
\label{11}\end{equation}
thus establish the contact with our previous momentum space
discussion \cite{2}.
\section{Feynman Diagrams and Distributions}
We shall set up rules for calculating
the singular expansion terms in (\ref{Fig 2}) which will yield
the following
values:
\begin{eqnarray}
&&\!\!\!\!\hspace{0mm}\raisebox{-1mm}{\mbox{\input 6.tps }} \!=
- \Delta(0)\,\Delta_{\mu \mu}(0)
 \mathop{=}_{D\rightarrow 1}
-
    \frac{1}{4},\!\!\!\!\!
\label{a1}
 \\
  &&
 \!\!\!\!\!\! \!\!\!\!\!\! \hspace{6mm}\raisebox{-3.2mm}{\mbox{\input 7.tps }}\! =
  - \Delta^2(0)\,\Delta_{\mu \mu}(0)
      \mathop{=}_{D\rightarrow 1}
 - \frac{1}{8m},
\label{a2}\end{eqnarray}
\begin{eqnarray}
\!\!\!\!\hspace{0mm}\raisebox{-1mm}{\mbox{\input 9.tps }} &\!\!\! = & \Delta^2(0)
\int d^Dx\, \Delta _{\mu \nu }^2(x)\!= \!
\Delta^2(0)
\int d^Dx\, \Delta _{\mu \mu }^2(x)
\nonumber\\   & = &
 - \left(1+\lfrac{D}{2}\right) m^2 \Delta^3(0)
\mathop{=}_{D\rightarrow 1}
    - \frac{3}{16m},
\!\!\!\!\!\!\!\!\!\label{b1}\\
\!\!\!\! \hspace{0mm}\raisebox{-1mm}{\mbox{\input 10.tps }} &\!\!\! = &
 \Delta_{\mu \mu}^2(0)
 \int d^Dx \,
 \Delta ^2 (x) =
 \left(1-\lfrac{D}{2}\right) m^2 \Delta^3(0)
 \nonumber\\  &\!\!\!\!
\displaystyle\mathop{=}_{D\rightarrow 1}&
        \frac{1}{16m},
\!\!\!\!\!\!\!\!\!\label{b2}\\
 \!\!\!\!\hspace{0mm}\raisebox{-1.5mm}{\mbox{\input 8.tps }}  &\!\!\! \!\!=&
  - \Delta(0)\Delta_{\mu \mu}(0)
  \int d^Dx\, \Delta _{\mu }^2(x)
 \nonumber\\  &\!\!\!\!\!= &
 -\left(\lfrac{D}{2}\right) m^2 \Delta^3(0)
\displaystyle\mathop{=}_{D\rightarrow 1}
 - \frac{1}{16m},
\!\!\!\!\!\!\!\!\!\label{b3}\end{eqnarray}
\begin{eqnarray}
\hspace{-4mm} \hspace{2mm}\raisebox{-2mm}{\mbox{\input 11.tps }} &=&
 \int d^D x\,  \Delta ^2 (x)  \Delta ^2_{\mu \nu }(x)
 \nonumber\\
 \hspace{-4mm} & = & \frac{m^2 }{ 3 (3D-4) } \left[ ( 8-7D) \Delta ^3 (0)
\phantom{\int}
 \right.\nonumber \\
 \hspace{-4mm}  & +&   \left.  (D+ 4) m^2 \int d^D x\,  \Delta ^4 (x) \right]
 \mathop{=}_{D\rightarrow 1} - \frac{3}{32m},
\label{rest16}\\
\hspace{-4mm}\!\!\!\!\!\!\!\! \hspace{2mm}\raisebox{-1.9mm}{\mbox{\input 12.tps }}   & = &
\int d^D x\,  \Delta (x)  \Delta _\mu(x)  \Delta _ \nu
    (x)  \Delta _{\mu \nu }  (x)  \nonumber \\
\hspace{-4mm}   & = &  - \frac{m^2}{6 (3 D - 4)} \left[ (5D-8)  \Delta ^3 (0)
\phantom{\int}
          \right. \nonumber \\
\hspace{-4mm}   & &  \left.  - 2  (D-4) m^2 \int d^D x\,  \Delta ^4 (x)
 \right]
     \mathop{=}_{D \rightarrow 1}  - \frac{1}{32m} ,
\label{rest17}\\
\hspace{-4mm}\hspace{2mm}\raisebox{-2mm}{\mbox{\input 13.tps }}  &= &
\int d^D x\,   \Delta ^2_\mu (x)  \Delta ^2_ \nu  (x)
 \nonumber \\
 \hspace{-4mm} & = & \frac{m^2}{3 (3D-4)} \left[ 2 (D-2)  \Delta ^3 (0)
\phantom{\int}
     \right. \nonumber\\
  \hspace{-4mm}  & +&   \left.   (D+4) m^2 \int d^D x\,  \Delta ^4 (x) \right]
       \mathop{=}_{D \rightarrow 1}   \frac{1}{32m}.
\label{rest18}\end{eqnarray}

 Summing up all contributions, we recover (\ref{exp1}), thus
confirming the invariance of the perturbatively defined
path integral under coordinate transformations.

\section{Bessel Function Representation}
In
$D=1- \varepsilon $ dimensions,
the distributions in the Feynman integrals
can be expressed
in terms of modified  Bessel functions $K_ \nu (z)$.
The basic correlation function
in $D$-dimension
is obtained from the
integral in
 Eq.~(\ref{1})
as
\begin{equation}
  \Delta (x) = c_D \, z^{1 - \lfrac{D}{2}} \, K_{1 - {D}/{2}}(z),
\label{rest2-1}\end{equation}
where $z \equiv m \, | x |$ is reduced length of
$ x_\mu$, with the usual euclidean norm
$|x| =  \sqrt{x_1^2 + \dots + x_D^2}$, and $K_{1 - \lfrac{D}{2}}(z)$
is the modified Bessel function. The constant factor in front is
$
 c_D = {m^{D-2}}/{(2 \pi ) ^{D/2} }.
$
 In one dimension, the correlation function
(\ref{rest2-1})
reduced to (\ref{@del1}). The short-distance
properties
of the correlation functions
is governed  by the small-$z$
behavior
of Bessel function at origin \cite{table}:
\begin{equation}
    K_ \nu (z)~ \mathop{\approx}_{z\approx 0}~
 \frac{1}{2}  \Gamma ( \nu ) (z/2)^{- \nu },~~~{\rm Re}~  \nu >0.
\label{rest2-3}\end{equation}
In the application
to path integrals,
we
set  the dimension equal to $D=1-\varepsilon$
with  a small positive
$\varepsilon$, whose limit
$ \varepsilon \rightarrow 0$ will
yield the desired
results in $D=1$ dimension.
In this regime,
Eq.~(\ref{rest2-3})
shows that the correlation function
(\ref{rest2-1}) is regular
at the origin:
\begin{equation}
  \Delta (0) = \frac{m^{D-2}}{(4 \pi )^{D/2}}  \Gamma
 \left(1 - \frac{D}{2}\right)\mathop{=}_{D=1}\frac{1}{2m}.
\label{rest2-4}\end{equation}

The first derivative of the correlation function (\ref{rest2-1}),
which is
the $D$-dimensional
extension of time derivative (\ref{@del2}), reads
\begin{equation}
   \Delta _\mu (x) = - c_D z^{1 - \lfrac{D}{2}} \,
 K_{\lfrac{D}{2}} (z) \, \partial _\mu z,
\label{rest2-5}\end{equation}
where $\partial_\mu z = m \, x_\mu/|x|$.
By  Eq.~(\ref{rest2-3}), this  is regular at the origin for $ \varepsilon>0$,
such that the
antisymmetry $ \Delta _\mu(-x) = -  \Delta _\mu (x)$ makes
$\Delta _\mu (0) =0 $.

In contrast to these two correlation functions, the second derivative
\begin{eqnarray}
 &&\Delta _{\mu \nu } (x) =  \Delta (x) \, (\partial_\mu z)
   (\partial_ \nu z)
 \!+\!\frac{c_D}{D\!-\!2}   z^{\lfrac{D}{2}} K_{\lfrac{D}{2}}
(z) \, \partial_{\mu \nu } z^{2-D}
\nonumber \\&&
\label{rest2-6}\end{eqnarray}
is singular at short distance. The singularity comes from
the second term in (\ref{rest2-6}):
\begin{equation}
 \partial _{\mu \nu } z ^{2 - D} = (2-D)\frac{m^{2-D}}{|x|^D}
    \left(\delta_{\mu \nu } - D \, \frac{x_\mu x_ \nu }{x^2}\right),
\label{rest2-7}\end{equation}
which is a distribution that is ambiguous at origin,
and defined up to the
addition of a $ \delta ^{(D)}(x)$-function. It is
regularized in the same way as the divergence
in the Fourier representation
Eq.~(\ref{11}). Contracting the indices $\mu$
and $ \nu $ in
Eq.~(\ref{rest2-7}),
we obtain
\begin{equation}
  \partial ^2 z^{2-D} = (2-D)m^{2-D} \, S_D \,
  \delta ^{(D)}(x),
\label{rest2-8}\end{equation}
where $ S_D =2  \pi ^{D/2} / \Gamma (D/2)$ is the surface of a sphere
in $D$ dimension. Contracted
Eq.~(\ref{rest2-6})
\begin{eqnarray}
  \!\Delta _{\mu\mu} (x) \!&=&\!
    m^2 \,  \Delta (x)\! -\! c_D \,
         m ^{2-D}  \, S_D \, \frac{1}{2} \,  \Gamma
    \left(\lfrac{D}{2}\right) 2 ^{D/2} \,
\delta ^{(D)}(x) \nonumber \\
    \!&=&\!
 m^2 \,  \Delta (x) -  \delta ^{(D)}  (x),
\label{15}\end{eqnarray}
coincides with the definition of the correlation function
by the inhomogeneous
field equation
\begin{equation}
(-\partial _\mu ^2+m^2)q (x)= \delta^{(D)} (x) .
\label{@eom}\end{equation}
From Eqs.~(\ref{11}) and (\ref{15}) we see that
\begin{equation}
  \Delta _{\mu\mu } (0) = m^2 \,  \Delta (0) \mathop{=}_{D\rightarrow 1}
     \frac{m}{2},
\label{16}\end{equation}
which determines the Feynman diagrams (\ref{a1}) and (\ref{a2}).

  A further
relation between distributions
is found
from the derivative
\begin{eqnarray}
&&\!\! \partial _\mu \,  \Delta _{\mu \nu }  (x) =
   \partial _ \nu  \left[ - \delta ^{(D)} (x) + m^2 \,  \Delta  (x)
   \right]  \nonumber \\
 &&+   m S_D \left[  \Delta (x) |x|^{D-1} (\partial _ \nu z)\right]
     \delta ^{(D)} (x) = \partial _ \nu   \Delta _{ \lambda  \lambda }
         (x).
\label{rest2-11}\end{eqnarray}
\comment{
Squaring Eq.(\ref{rest2-6}) gives
\begin{eqnarray}
 &&\Delta ^2 _{\mu \nu }  (x) = m^4 \,  \Delta ^2 (x) - m^4 \,
         c_D^2 (D-1)z^{1-D} \frac{d}{dz} \left[K^2_{\lfrac{D}{2}}(z)\right]
 \nonumber \\
& & - m^{4-D} \, c_D^2 \,  \Gamma  \left(\lfrac{D}{2}\right)
      \Gamma  \left(1 - \lfrac{D}{2}\right) S_D \,  \delta ^{(D)} (x)
 \nonumber \\
 &&
  - (D-1) m^{4-2D} \, c_D^2 \, 2 ^{D-1} \,  \Gamma ^2 \left(
    \lfrac{D}{2}\right) S_D^2 \, \left[ \delta ^{(D)} (x) \right]^2 .
\label{rest2-12}\end{eqnarray}
}
   Let us now turn to the calculation of Feynman
integrals (\ref{b1})--(\ref{rest18}) over products
of distributions.

\section{Integrals over Two Distributions}
Consider now the integrals over
products of two such distributions. If an integrand
$f(|x|)$
depends only on $|x|$,
we may perform the integrals over
the directions of the vectors
\begin{equation}
 \int d^D x\, f(x) ={S_D} \int^{\infty}_{0} dr \,
   r^{D-1} \, f(r),~~~~~r\equiv |x|.
\label{rest2-13}\end{equation}
Thus we can  calculate directly \cite{table}:
\begin{eqnarray}
  &&\!\!\!\!\!\!\!\!\!\! \int d^D x\, \,  \Delta ^2 (x) = m ^{-D} \, c_D^2 \,
       S_D \int^{\infty}_{0} dz \, z \, K^2_{1 - \lfrac{D}{2}}(z)
 \nonumber \\
&& = m^{-D} \, c_D^2 \, S_D \frac{1}{2} \left(1 - \lfrac{D}{2}\right)
            \Gamma \left(1 - \lfrac{D}{2}\right)  \Gamma
       \left(\lfrac{D}{2}\right) \nonumber \\
&& = \frac{2-D}{2m^2}   \Delta (0).
\label{rest2-14}\end{eqnarray}
and
\begin{eqnarray}
&&  \int d^D x\,  \Delta ^2 _\mu (x) = m^{2-D}\, c_D^2 \, S_D
 \, \int^{\infty}_{0} dz \, z K^2 _{\lfrac{D}{2}} (z)  \nonumber \\
&&
 ~=m^{2-D}  c_D^2 S_D  \frac{1}{2}  \Gamma \left(1 +
 \lfrac{D}{2} \right)  \Gamma  \left(1 - \lfrac{D}{2}\right) =
     \frac{D}{2}  \Delta (0),
\label{rest2-15}\end{eqnarray}
which together with (\ref{16}) explains the values given for
the Feynman integrals in Eqs.~(\ref{b2}) and (\ref{b3}).

 Note that due to the relation \cite{table}:
\begin{equation}
   K_{\lfrac{D}{2}} (z) = -z^{\lfrac{D}{2} -1} \,
    \frac{d}{dz} \left[ z^{1 - \lfrac{D}{2}} \, K_{1- \lfrac{D}{2}}(z)
    \right] ,
\label{rest2-16}\end{equation}
the integral over $z$ in Eq.~(\ref{rest2-15}) can also be performed
by parts,
yielding
\begin{eqnarray}
 &&\!\int d^D x\,  \Delta _\mu^2 (x)\nonumber \\
&& = -m^{2-D} \, c_D^2
    \, S_D \left(z^{\lfrac{D}{2}} K_{\lfrac{D}{2}} \right) \left(
     z^{1 - \lfrac{D}{2}} K_{1 - \lfrac{D}{2}}\right)
    \bigg\vert_{0}^{\infty}  \nonumber \\
&& ~~~- m^2  \int d^D x  \Delta ^2 (x) =
 \Delta (0) -m^2 \int d^D x\,  \Delta ^2 (x) .
\label{rest2-17}\end{eqnarray}
The upper limit on the right-hand side
does not contribute because of the
exponentially fast decrease of the Bessel function at infinity
\cite {table}.

Using the explicit representation
(\ref{rest2-6}), we calculate similarly the integral
\begin{eqnarray}
&& \!\!\!\!\!\!\!\int d^D x\,  \Delta ^2 _{\nu \nu } (x)
= \int d^D x\,
\Delta ^2 _{ \mu\nu }
  \label{rest2-19}\label{21} \\
&=& m^4 \int d^D x\,  \Delta ^2 (x)
 - m^{4-D} \, c_D^2 \,  \Gamma \left(\lfrac{D}{2}\right)
  \Gamma \left(1 - \lfrac{D}{2}\right) S_D   \nonumber \\
&=& m^4 \, \int
    d^D x\,  \Delta ^2 (x) - 2m^2 \,  \Delta (0)=
-
  \left(1+\lfrac{D}{2}\right) m^2 \Delta (0)\,.
\nonumber\end{eqnarray}
The first equality follows from  two partial integrations
and Eq.~(\ref{rest2-11}).
In the last equality
we have used  (\ref{rest2-14}).
We have omitted an integral
containing modified Bessel functions
\begin{eqnarray}
&& \!\!\!\!\!\!\!\!\! (D-1) \, \left[ \int ^{\infty}_{0} dz K_{\lfrac{D}{2}}
 (z) K_{1 - \lfrac{D}{2}} (z)\, + \right. \nonumber \\
&&  \left. ~~~~~~~~~~~~~
 + \frac{D}{2} \int^{\infty}_{0} dz \, z^{-1}
    \, K^2_{\lfrac{D}{2}} (z)\right]
\label{rest2-20a}\end{eqnarray}
since this vanishes
in one dimensions as follows:
\begin{eqnarray}
 - \frac{ \pi }{4}   \Gamma \left(1- \lfrac{\varepsilon}{2}\right)
    \left[  \Gamma \left(\lfrac{\varepsilon}{2}\right)+  \Gamma
    \left(- \lfrac{\varepsilon}{2}\right)\right] \,
   \varepsilon^2 \,  \Gamma (\varepsilon) \mathop{=}_{\varepsilon
    \rightarrow 0} 0. \nonumber
\label{rest2-20}\end{eqnarray}
The result (\ref{21})
explains the
value     stated
for the Feynman integral in Eq.~(\ref{b1}).

The results
(\ref{rest2-14}), (\ref{rest2-15}),
(\ref{21}) can be used to derive a fundamental
rule that the integral
over the square of the $ \delta $-function vanishes.
Indeed, solving the inhomogenous field equation
(\ref{15})  for $  \delta ^{(D)} (x)$,
and squaring it, we obtain
\begin{eqnarray}
  &&\!\!\!\!\!\!\!\!\!\!\int d^D x\,\,\left[\delta^{D} (x)\right]^2 =
 m^4 \int d^D x\, \, \Delta^2 (x)\nonumber\\
&& +
 2m^2 \int d^D x\, \,\Delta^2_{\mu} (x) +
 \int d^D x\, \, \Delta^2_{\nu \nu} (x) = 0.
\label{id}\end{eqnarray}
Thus we may formally calculate
\begin{eqnarray}
\int d^D x\,\,
\delta^{D} (x)
\delta^{D} (x)
=\delta^{D} (0)=0,
\label{@}\end{eqnarray}
pretending that one of the two $ \delta $-functions
is an admissible
 test function of ordinary distribution theory.

\section{Integrals over  Four Distributions}
We now  calculate the
integrals over products of four
distributions required for the diagrams
(\ref{rest16})--(\ref{rest18}).
These integrals
are straightforward
in $D=1$ dimension,
as long as they are unique.
Only
ambiguous cases require a calculation in
$D = 1-\varepsilon$ dimension,
with the limit $\varepsilon \rightarrow 0$
taken at the end.

A unique case is
\begin{eqnarray}
 &&\!\!\!\!\!\!\!\!\!\!\!\!\!\!\!\!\!\int d^D x\,  \Delta ^4 (x) = c_D^4 \, m^{-D} S_D
 \int^{\infty}_{0} dz \, z^{3-D} \,
 K_{1-\lfrac{D}{2}}^4 (z)
 \nonumber \\
~~~~\mathop{\approx}_{ D\approx1}
 & &  c_1^4 \, m^{-1} \, S_1 \frac{ \pi^2}{2^4}
  \Gamma ^4 \left(\frac{3}{2} - \frac{D}{2} \right) \Gamma (D)
\mathop{=}_{ D\rightarrow 1}
    \frac{1}{32 m^5}.
\label{rest2-21}\end{eqnarray}
Similarly, we calculate
\begin{eqnarray}
\lefteqn{\!\!\!\!\!\!\!\!\!\!
 \int d^D x\,  \Delta ^2 (x)  \Delta ^2 _\mu (x) } \nonumber \\
 & &= m^{2-D}
     \, c_D^4 \, S_D \int_{0}^{\infty} dz z^{3-D}
    \, K^2_{\lfrac{D}{2}} (z) K^2_{1 - \lfrac{D}{2}} (z)
 \nonumber \\
 & & = \frac{1}{3} m^{2-D} \, c_D^4  \, S_D \, \left.\bigg[
   2^{-D-1} \,  \Gamma \left(\lfrac{D}{2}\right)  \Gamma ^3
    \left(1 - \lfrac{D}{2}\right)  \right.  \nonumber \\
& &  \left. +\int^{\infty}_{0} dz  \left(z^{1 - \lfrac{D}{2}} \,
   K_{1 - \lfrac{D}{2}}\right)^3 \frac{d}{dz} \left(z^{\lfrac{D}{2}}
  \, K_{\lfrac{D}{2}}\right)\right]  \nonumber \\
& & =\frac{1}{3} \left[  \Delta ^3  (0) - m^2 \int
   d^D x\,  \Delta ^4 (x) \right.\bigg]  \mathop{=}_{ D\rightarrow 1}
 \frac{1}{32m^3}.
   \label{rest2-22}\end{eqnarray}
Using the expressions~(\ref{rest2-5})
and (\ref{rest2-6}), we find
for  the
integral in $D = 1- \varepsilon$ dimensions
\begin{eqnarray}
 & & \!\!\!\!\!\!\!\!\!\!\!\!\!\!\!\!
\!\!\!\!\!\!\int d^D x\,  \Delta  (x)  \Delta _\mu (x)  \Delta _ \nu  (x)
       \Delta _{\mu \nu } (x)  \nonumber \\
 & & =m^2 \, \int d^D x\,  \Delta ^2 (x)  \Delta ^2 _\mu
 (x) + I_D,
\label{rest2-23}\end{eqnarray}
where  $I_D$
denotes the singular
integral
\begin{eqnarray}
   I_D & = & (D-1) m^{4-D} \, c^4_D  \, S_D  \,
  \nonumber \\ & & \times
\int^{\infty}_{0} dz z^{2-D}  \, K_{1- \lfrac{D}{2} }(z)
  K^3 _{\lfrac{D}{2}} (z).
\label{rest2-24}\end{eqnarray}
In spite of the prefactor $D-1$,
this
has a nontrivial limit for
$D\rightarrow 1$,
the zero being
 compensated by a  pole from the small-$z$ part of the integral.
at $z=0$.
In order to see this we
use
the  integral representation of the Bessel function \cite{cit}
\begin{eqnarray}
K_ \nu  (z) & = &   \pi^{-1/2} \, (z/2)^{- \nu } \,
     \Gamma \left(\frac{1}{2} +  \nu \right)
\nonumber \\
&&\times  \int^{\infty}_{0}
 dt (\cosh t)^{-2 \nu } \, \cos (z \sinh t ).
\label{rest2-29}\end{eqnarray}
In one dimension where $ \nu =1/2$, this becomes simply
$K_{1/2}(z) = \sqrt{\lfrac{\pi}{2z}} e^{-z}$.
For $\nu=D/2$ and $\nu=1-D/2$ written as
$ \nu =(1\mp\varepsilon)/2$,
it is approximately equal to
\begin{eqnarray}
&&\!\!\!\!\!\!K_{(1\mp\varepsilon)/2}(z)  =  \pi^{-1/2} \, (z/2)^{-(1\mp\varepsilon)/2 } \,
 \Gamma \left(1\mp\frac{\varepsilon}{2} \right)\label{as} \\
&\times&
\left[ \frac{\pi}{2} e^{-z} \pm \varepsilon
  \int^{\infty}_{0}
   dt (\cosh t)^{-1} \,\ln (\cosh t)\, \cos (z \sinh t ) \right] ,\nonumber
\end{eqnarray}
where the $t$-integral is regular at $z=0$ \cite{cite2}.
After substituting (\ref{as}) into (\ref{rest2-24}),
we obtain the finite value
\begin{eqnarray}
I_D &\displaystyle \mathop{\approx}_{\varepsilon\approx0} & -\left(m^{4-D}  \, c_D^4 \, S_D\right)
 \varepsilon \, \nonumber \\
 & &\times  \frac{ \pi ^2}{4}  \Gamma \left(1 + \lfrac{\varepsilon}{2}
  \right) \Gamma ^3 \left(1 - \lfrac{\varepsilon}{2}\right) \times
2^{-5\varepsilon} \,  \Gamma  (2 \varepsilon)
   \nonumber \\
 & \displaystyle \mathop{=}_{\varepsilon\rightarrow 0}&- \left(\frac{1}{2m\pi^2}\right)\,
 \frac{ \pi ^2}{8} = - \frac{1}{16m}.
\label{rest2-30}\end{eqnarray}
The prefactor $ D-1=-\varepsilon $ in
(\ref{rest2-24}) has been canceled by
the pole in $\Gamma (2\varepsilon)$.

The nontrivial nature of the integral $I_D$
 was first
observed in another form
in the momentum space calculations
of Ref.~\cite{2},
where $I_D$ appeared
in the integral
\begin{equation}
  \int d^D x\,  \Delta ^2 (x) \left[  \Delta ^2_{\mu \nu } (x)
        -  \Delta ^2_{ \lambda  \lambda } (x) \right] =
      - 2 I_D,
\label{rest2-28}\end{equation}
Indeed, using the Bessel expressions
(\ref{rest2-5}) and
(\ref{15}), we find
\begin{eqnarray}
 && \int d^D x\, \,  \Delta ^2 (x) \left[  \Delta ^2 _{\mu \nu }
 (x) -  \Delta ^2_{ \lambda  \lambda } (x) \right]  = - (D-1) m^{4-D}
\nonumber  \\
 && ~~~~~\times \,
 c_D^4  S_D
     \int^{\infty}_{0} dz \left(z^{1 - \lfrac{D}{2}} \,
   K_{1 - \lfrac{D}{2}}\right)^2 \frac{d}{dz}  K^2_{\lfrac{D}{2}},
\label{rest2-26}\end{eqnarray}
and a partial integration
\begin{eqnarray}
\lefteqn{ \!\!\!\!\!\!\!\!\!\!\! \int_{0}^{\infty} dz \left(z^{1 - \lfrac{D}{2}} \,
         K_{1- \lfrac{D}{2}}\right)^2 \frac{d}{dz} K^2_{\lfrac{D}{2}}
} \nonumber \\ &&~~~~~~~=
 2\int_{0}^{\infty} dz \, z^{2-D}  \, K_{1 - \lfrac{D}{2}}
    \, K^3_{\lfrac{D}{2}}
 \label{rest2-27}\end{eqnarray}
establishes contact with the integral
(\ref{rest2-24}) for
$I_D$.

Knowing $I_D$, we also determine,
after integrations by parts,
the integral
\begin{equation}
  \int d^D x\,  \Delta ^2_\mu (x)  \Delta ^2_ \nu  (x) = -
  3m^2 \int d^D x\,  \Delta ^2 (x)  \Delta ^2_\mu (x) -
 2 \, I_D.
\label{rest2-25}\end{equation}

It remains to calculate one more unproblematic integral over four distributions:
\begin{eqnarray} \label{rest2-31}
& &\!\!\!\!\!\! \int d^D x\,  \Delta ^2 (x)  \Delta ^2_{ \lambda  \lambda }(x)
      \\
 & &= \left[ - 2m^2 \,  \Delta ^3 (0) + m^4 \int d^D x\,  \Delta ^4 (x)
 \right] _{ D=1} = - \frac{7}{32m}.   \nonumber
\end{eqnarray}
Combining this with
(\ref{rest2-30}) and (\ref{rest2-25}) we find the
the Feynman diagram (\ref{rest16}).
The combination of (\ref{rest2-23}) and (\ref{rest2-25})
with (\ref{rest2-30}) and (\ref{rest2-22}), finally, yields
the diagrams (\ref{rest17}) and  (\ref{rest18}), respectively.

\section{Summary}
In this note we have shown that
with appropriate Bessel function representations,
we can evaluate
integrals over products of distributions in configuration space
which
reproduce the results
of
dimensional regularization.
This ensures
the invariance of
perturbatively defined  path integral
under coordinate transformations observed in \cite{1}.

\end{document}